\documentclass{PoS}

\title{Generalized fluxes in matrix  compactifications}

\ShortTitle{Matrix flux compactifications}

\author{\speaker{Athanasios Chatzistavrakidis}
\\
        Institut f\"ur Theoretische Physik, Leibniz Universit\"at Hannover, \\
Appelstra{\ss}e 2, 30167 Hannover, Germany\\
        E-mail: \email{thanasis@itp.uni-hannover.de}}

\author{Larisa Jonke\\
        Theoretical Physics Division, Rudjer Bo\v skovi\'c Institute, \\ Bijeni\v cka 54, 10000  Zagreb, Croatia\\
        E-mail: \email{larisa@irb.hr}}

\abstract
{In the recent years a lot of attention is focused on unconventional string compactifications. A variety 
of different but related frameworks was developed in order to address issues such as duality invariance, non-geometry 
and non-commutativity in string theory. In this contribution we review and clarify the approach that 
goes through matrix models. Furthermore, we discuss some connections of this framework to other related approaches.}

\FullConference{Proceedings of the Corfu Summer Institute 2012 "School and Workshops on Elementary Particle Physics and Gravity"\\
		September 8-27, 2012\\
		Corfu, Greece}

\newcommand{\sfrac}[2]{{\textstyle\frac{#1}{#2}}}

\newcommand{\be}{\begin{equation}}
\newcommand{\ee}{\end{equation}}

\newcommand{\beq}{\begin{equation}}

\newcommand{\eeq}{\end{equation}}

\newcommand{\beqa}{\begin{eqnarray}}

\newcommand{\eeqa}{\end{eqnarray}} 

\def\nn{\nonumber} \def \bea{\begin{eqnarray}} \def\eea{\end{eqnarray}}

\newcommand{\barr}{\begin{array}}

\newcommand{\earr}{\end{array}}

  \def\G{\Gamma}

 \def\d{\delta} 

 \def\e{\epsilon}

 \def\L{\Lambda}

\def\Z{{\mathbb Z}} \def\one{\mbox{1 \kern-.59em {\rm l}}}

\def\bit{\begin{itemize}} \def\eit{\end{itemize}}

\def\({\left(} \def\){\right)}

\begin{document}

\section{Introduction and discussion}

The low-energy limit of perturbative string theories is described by particular ten-dimensional supergravities, which 
are very useful in the study of string compactifications. However, unlike supergravity, string theory is not a field theory. 
Therefore it has properties, mainly related to the finite string length,
which cannot be directly captured by vanilla low-energy supergravity. Such properties include the existence of winding modes 
of the string, string dualities, non-geometry and relations to non-commutative geometry to name a few. Their role in 
string compactifications is under investigation, which led to the development of several new frameworks and techniques. 
Indeed, the above issues were addressed systematically in the doubled formalism \cite{hullgeomfornongeom} and in 
particular with the construction of 
T-folds and twisted doubled tori \cite{hulltdt,prezastdt}. Moreover, techniques from  generalised complex geometry 
\cite{gg2} have also proven to be useful in this 
context \cite{Grana:2008yw,Berman:2010is,Andriot1}. A powerful framework where both the doubled formalism and 
generalised geometry come into play is 
double field theory \cite{dft}. However, it should be mentioned that the above frameworks, albeit capturing 
more properties of the fundamental string theory than vanilla supergravity does, 
are effective theories too. At a more fundamental level, use of 
conformal field theory methods and sigma models have also provided interesting insights \cite{Lust1,BP,CFL,szabong}. 
Here we would like to argue that a fundamental framework where unconventional string compactifications can be studied 
is matrix models. Such an approach was initially pursued in Ref. \cite{ramgoolam} and more recently in Refs.
\cite{ChatzistavrakidisJonke1,ChatzistavrakidisJonke2}. 

The motivation to use matrix models in order to study unconventional string compactifications is manifold. 
Certain matrix models, 
with prime examples the BFSS and IKKT models \cite{Banks:1996vh,Ishibashi:1996xs}, grew out from efforts to understand 
second quantization of strings and 
there is evidence that they can be considered as non-perturbative definitions of M theory or superstrings (some reviews 
on the topic are Refs. \cite{Banks:1997mn,Bigatti:1997jy,Taylor:1997dy}). It is advantageous that such models are 
non-perturbative, quantum theories where non-commutative 
structures naturally appear  \cite{Szaboreview}. Moreover, they are more than abstract constructions since they may be actually utilized 
for model building in particle physics and cosmology \cite{Aoki3,Chatzistavrakidisinter,Kim:2012mw}
. Regarding compactifications, there are essentially two points 
of view. One way would be to directly define the matrix model on a compact manifold, 
while the other one involves imposing conditions
to the degrees of freedom corresponding to different compactifications of the matrix model
{\footnote{It is worth mentioning  another approach which is different in spirit than traditional string 
compactifications \cite{Steinacker:2011wb}. In this approach spacetime structures are viewed as compact brane-like solutions 
of matrix models embedded in flat ten-dimensional ambient spacetime.}}. This second point of 
view was employed in Ref. \cite{cds}, where toroidal matrix compactifications were studied and it was shown that there exist 
many interesting connections to non-commutative geometry. In particular, compactifications of matrix theory on non-commutative tori seem to 
point at a correspondence between deformation parameters of the tori and moduli of 11D supergravity 
\cite{cds,Douglas:1997fm,bmz}. Then it is 
reasonable to ask whether such a correspondence is relevant for compactifications with fluxes. Fluxes in string 
compactifications can be either purely geometric, related to the internal geometry of the compactification manifold, 
NSNS or RR fluxes, being expectation values of form fields, or the so-called non-geometric fluxes. The latter are 
dual cousins of the previous ones and they have been discussed extensively in the literature (see Ref. \cite{Shelton:2005cf} 
and its citations). The purpose here 
is to review and clarify how such quantities can be traced in matrix compactifications. 

A convenient starting point is to determine and solve the conditions which define a matrix compactification on 
nilmanifolds. The latter are smooth parallelizable manifolds with non-trivial geometry which incorporate geometric 
fluxes and they have played a distinguished role in string compactifications 
\cite{Scherk,km,kstt,Hulltts,prezasg2,Hullttm,Grana2,Caviezel:2008ik}. Their study in the matrix model is 
tractable because they can be described as non-trivial toroidal fibrations. This picture emerges from their description 
as homogeneous spaces constructed as orbits of a lattice in a nilpotent Lie group {\footnote{Relations between 
nilpotent Lie algebras and matrix models
were studied systematically in Ref. \cite{Chatzistavrakidis:2011su}.}}.
The prototype example of such a 
construction is the 3D Heisenberg nilmanifold but there exist  several more cases with richer geometry. This 
construction leads to a set of identification conditions,  generalizing the ones of a standard torus, which can 
subsequently be imposed on the degrees of freedom of the matrix model in the same spirit as for the toroidal matrix 
compactifications of Ref. \cite{cds}. Associated to 
matrix compactifications on nilmanifolds, one obtains particular operator algebras which  can be represented in terms of phase space variables. In the T-dual 
picture one obtains a non-commutative Yang-Mills theory on a dual nilmanifold, as usual in matrix compactifications
\cite{cds,bmz,Ho:1997yk}.

Matrix compactifications on tori and nilmanifolds show that  an important role in this context is played by phase space, 
as expected in a quantum-mechanical theory. Apart from the original degrees of freedom $X$, new objects $\tilde{X}$
appear, a situation reminiscent of the appearance of winding modes when strings are compactified. This natural doubling 
can subsequently be used to examine other duality frames in matrix compactifications. Indeed, we argue that 
particular transformations lead to matrix compactifications which can be identified with situations where NSNS flux 
or non-geometric fluxes are present. We present  two arguments in favour of this interpretation. The first is 
that the action of the operator algebras that we obtain  on the generalized phase space of $(X, \tilde{X})$ exhibits 
the structure of a class of twisted doubled tori as constructed in Refs. \cite{hulltdt,prezastdt} in order to account for the 
corresponding geometric and non-geometric fluxes. Let us stress that we do not impose the conditions of these twisted doubled tori 
but they rather follow from the action of the operator algebras. Secondly, the formulae that provide the 
flux in each matrix compactification as derivation of the commutator of the matrices are in exact correspondence with 
the ones found for the same fluxes in the context of double field theory \cite{DF1,DF2,Andriot2,Geissbuhler:2013uka}. 
Therefore, each operator algebra is associated to some matrix compactification and each matrix compactification 
corresponds to a duality frame of a flux background in supergravity. This provides a picture of duality frames as different 
superselection sectors of the matrix model, accompanied with flux quantization which is part of the consistency of the compactification.

According to the above, some understanding of unconventional flux compactifications of string theory is possible within the 
framework of matrix models. Moreover, this formalism bears a lot of resemblance and is in agreement to other approaches. It is then reasonable to ask what more could we learn following this road that we could not learn 
in another way. This critical viewpoint brings us partially back to our motivation and partially to a discussion on 
future possibilities. From a motivational point of view, as already mentioned before, matrix models are not supposed to be 
effective theories but rather constructive definitions of string or M theory. If this is true, then understanding 
properties of string compactifications in this framework would be more valuable than in an effective theory. However,  matrix models still need to prove their merit. A direction which we find interesting is to 
make an effort to describe genuinely non-geometric cases in the matrix model. What we mean here is that the non-geometric 
cases that are usually discussed possess duality frames where they become geometric. Since T-dual backgrounds are 
physically equivalent this means that they are not really new. What would be totally new is a non-geometry which cannot be 
geometrized in any duality frame (see the discussion in Ref. \cite{Dibitetto:2012rk}). This issue was also discussed in Ref. \cite{deBoer:2012ma} in the context of general co-dimension 2 
objects dubbed exotic branes. It would also be interesting to study the dynamics of such objects and  matrix models 
might play a role in this too.

\section{Matrix compactification on the Heisenberg nilmanifold}

\subsection{Geometry of the unpolarized 3D nilmanifold}

Let us examine the 3D nilmanifold based on the nilpotent Lie algebra $\mathfrak g=(0,0,12)$ with corresponding group $G$.
This notation means 
that the algebra has three generators $T_i,~i=1,2,3$ which satisfy the single commutation relation $[T_1,T_2]=T_3$. 
The only non-vanishing structure constants are therefore $f_{12}^3=-f_{21}^3=1$. The Heisenberg nilmanifold is obtained 
as an orbit of a lattice $\G\subset G$ in $G$. It should be noted that there is some freedom on the construction of such 
an orbit. The most usual choice in the physics literature is the polarised one, where 
the nilmanifold takes the form of a 2-torus fibration over a circle. Here we consider instead a different 
parametrization, which is 
more convenient when abstract index notation is used. 

Let us consider the following basis of left-invariant 1-forms,
\begin{eqnarray}
 e^1 &=& dx^1, \nn\\
 e^2 &=& dx^2, \nn\\
 e^3 &=& dx^3-\sfrac{1}{2}x^1dx^2+\sfrac{1}{2}x^2dx^1,
\end{eqnarray}
which satisfy the Maurer-Cartan equations
$$de^i+\sfrac 12 f^i_{jk}e^j\wedge e^k=0.$$
There is a basis of dual vector fields on the tangent bundle of the nilmanifold, which is found to be
\begin{eqnarray}
 e_1 &=& \partial_1-\sfrac{1}{2}x^2\partial_3, \nn\\
 e_2 &=& \partial_2+\sfrac{1}{2}x^1\partial_3, \nn\\
 e_3 &=& \partial_3.\label{vf}
\end{eqnarray}
One can directly confirm that 
$$\langle e^i,e_j\rangle=\d^i_j, $$
where $\langle\cdot,\cdot\rangle$ is the standard pairing between the algebra $\mathfrak g$ and its dual algebra 
$\mathfrak g^*$. These 1-forms and vectors are defined on the corresponding group manifold $G$, which is the covering 
space of the nilmanifold, and they descend on the orbit $\Gamma \backslash G$ under the 
following identifications:
\be
(x^1,x^2,x^3,\partial_1,\partial_2,\partial_3)\sim (x^1+c^1,x^2+c^2,x^3+\sfrac{c^1}{2}x^2-\sfrac{c^2}{2}x^1+c^3,
\partial_1+\sfrac{c^2}{2}\partial_3,\partial_2-\sfrac{c^1}{2}\partial_3,\partial_3). \ee
Note that we included the transformations of the derivative operators which follow from eq. (\ref{vf}). 
The above relation may as well be written in the following very compact form:
\be\label{ids}(x^i,\partial_j)
\sim (x^i+c^i+\sfrac{1}{2}f^i_{jk}c^jx^k,\partial_i+\sfrac{1}{2}f_{ij}^kc^j\partial_k).
\ee

\subsection{Matrix compactification}

We consider a set of matrices $X^a$ of a matrix model associated to them, for example the BFSS or the IKKT model. 
In the formulation employed here the only difference is the number of the matrices; the index $a$ runs from 1 to 
9 for the BFSS model, where time is treated separately, and from 1 to 10 for the IKKT model. Moreover, we focus 
our attention to the bosonic sector. Fermions $\psi$, being the 
supersymmetric partners of $X^a$ in each model, are treated similarly as discussed for example in Ref. \cite{cds}. 
The classical action of the BFSS model is 
\be\label{BFSSaction}
{ S}=\frac 1{2g}\int dt\biggl[ Tr\big(\dot  { X}^a\dot { X}_a
-\frac 12 [{X}^a,{X}^b]^2\big)
+2\psi^T\dot\psi-2\psi^T\Gamma_a[\psi,{X}^a]\biggl], 
\ee  
where $\G_a$ furnish a representation of $SO(9)$. This action is written in units of the string length and 
$g$ is the string coupling \cite{Banks:1996vh}.  

The most standard approach to toroidal compactification of a matrix model involves imposing a set of conditions which 
correspond to the identifications of the coordinates of the torus under the lattice action. This approach is also 
appropriate for compactifications on nilmanifolds, as shown in 
\cite{ramgoolam,ChatzistavrakidisJonke1,ChatzistavrakidisJonke2}.
In the present case it amounts to considering three unitary operators $U_i$ and the following set of conditions, 
corresponding to the identifications (\ref{ids}),
\bea \label{conditionstt3}
U_iX^iU_{-i}&=& X^i+R^i, \quad i=1,2,3,\nn\\
U_1 X^3U_{-1}&=& X^3+\sfrac {R^1}2 X^2 , \nn\\
U_2 X^3U_{-2}&=& X^3-\sfrac {R^2}2 X^1, \nn\\
U_i X^aU_{-i}&=& X^a, \quad a\ne i, \quad a=1,\dots,9,  \quad (a,i)\ne \{(3,1),(3,2)\}.
\eea
Note that we adopt the notation $U_{-i}\equiv U_i^{-1}$. $R^i$ are constant matrices which may be thought of as radii 
(more accurately, as radii multiplied by $2\pi$){\footnote{The reader should avoid confusion of indices and powers. The 
superscript in $R^2$ is an index and not a power. In this paper there will never appear a power of such a quantity, therefore 
one should always think of superscripts as indices.}.
Let us define the $Q$-operator as 
\be 
Q_{ij}=U_iU_jU_{-i}U_{-j}.
\ee
In the present case we find the following commutation relations:
\bea \label{qq}
[Q_{ij}, X^k]=-R^iR^jf_{ij}^{\ k}Q_{ij},
\eea
where no summation is implied on the RHS. Now, these relations imply that $Q_{13}$ and $Q_{23}$ are scalar operators and 
therefore we may write 
\be
Q_{13}=e^{-i\theta_{13}} \quad \mbox{and} \quad 
Q_{23}=e^{-i\theta_{23}},
\ee
for some real parameters $\theta_{13},\theta_{23}$.  
The third $Q$-operator, $Q_{12}$, does not commute with $ X^3$ and therefore we cannot directly write its form as for 
the previous two. Notice, however, that the relation (\ref{qq}) can be rewritten as
$$Q_{12} X^3Q_{-12}=X^3-R^1R^2f_{12}^{\ 3}.$$
Comparing this expression with conditions (\ref{conditionstt3}) we conclude that
\be \label{q12}
Q_{12}\propto U_{-3}.
\ee
Let us set all constants $R^i=1$ for the moment. 
Now, in order to solve the conditions (\ref{conditionstt3}), we write the unitary operators in a rather generic form $U_i=e^{i{ Y}_i}$, where $ Y_i$ are arbitrary 
hermitean operators. Note that the unitary operators act on functions of a Hilbert 
space and their form is specified by this action; this does not directly 
mean that the operators $ Y_i$ can be written down explicitly. However, 
in the present case it turns out that this is indeed possible. 
It is directly observed that the algebra of $U$s takes the following form,
\bea 
U_1U_3&=&e^{-i\theta_{13}}U_3U_1, \nonumber\\
U_2U_3&=&e^{-i\theta_{23}}U_3U_2, \nonumber\\
U_2U_1&=&U_3U_1U_2.
\eea
Let us mention that this is a $C^*$ algebra known in the mathematical literature from the work of Packer \cite{packer}. 
It is now simple to determine the algebra of $( X, Y)$. 
The ``diagonal'' compactification conditions imply that
$$[{ Y}_i, X^i]=-i.$$
Moreover, the ``off-diagonal`` ones imply that 
$$[{ Y}_1, X^3]=-\sfrac i2  X^2 \quad \mbox{and} \quad [{ Y}_2, X^3]=\sfrac i2  X^1.$$
Furthermore, Eq. (\ref{q12}) gives
$$[{ Y}_1,{ Y}_2]=i{ Y}_3.$$
Similarly, we obtain
$$[{ Y}_1,{ Y}_3]=i\theta_{13} \quad \mbox{and} \quad [{ Y}_2,{ Y}_3]=i\theta_{23}.$$
This procedure does not fix the commutation relations among $ X^i$.
Here we restrict to trivial gauge bundles  and thus the full algebra of 
$X$ and ${ Y}$ reads as 
\bea 
[ X^i,X^j]&=&0, \nonumber\\
{[} X^i,{ Y}_j]&=&i\delta_i^j+\sfrac i2 f_{jk}^{\ i} X^k,\nonumber\\
{[}{ Y}_i,{ Y}_j]&=&if_{ij}^{\ k}{ Y}_k+i\theta_{ij}|_{\not 12}, 
\eea
where $f_{12}^{\ 3}=1$ and the notation $|_{\not 12}$ means that $\theta_{12}$ vanishes{\footnote{The reason that
$\theta_{13}$ and $\theta_{23}$
are non-vanishing but $\theta_{12}$ vanishes may be formulated in terms of Poisson geometry. Indeed, Rieffel in Ref.
\cite{Rieffel}
shows that the most general invariant Poisson structure on the Heisenberg nilmanifold is $\L=\mu \partial_1\wedge \partial_3
+\nu \partial_2\wedge \partial_3$, where $\mu,\nu$ are constants. Quantization of this Poisson structure would then lead 
to non-vanishing commutators only for the corresponding pairs of indices.}}.

One question that arises is whether an explicit expression for the solutions may be found in terms of coordinate and 
momentum operators satisfying the standard Heisenberg relation
$$[\hat x^i,\hat p_j]=i\d^i_j$$
but otherwise commuting, i.e.
$$ [\hat x^i,\hat x^j]=0 \quad \mbox{and} \quad [\hat p_i,\hat p_j]=0.$$
This is indeed possible. For simplicity, let us consider the case of $\theta_{ij}=0$. Then a particular solution may 
be written as 
\bea \label{yy}
 X^i&=&\hat x^i, \nn \\
{ Y}_i &=& \hat p_i+\sfrac{1}{2}f_{ij}^{\ k}\hat x^j\hat p_k.
\eea
Note that the ${ Y}$s are related to the right-invariant vector fields on the nilmanifold.

As shown in Ref. \cite{cds}, having found a 
particular solution, the general solution is given as 
\be \label{gss}
 X^i = \hat x^i + \hat { A}^i,
\ee
where $\hat { A}^i=\hat { A}^i(\hat U_j)$ depend \footnote{Imposing compactification conditions (\ref{conditionstt3}) on 
the general solution (\ref{gss}) implies actually that 
$\hat { A}^i= { A}^i(\hat U_j)+\sfrac{1}{2}f_{jk}^{\ i}\hat x^j{ A}^k(\hat U_j)$, see Ref. \cite{ChatzistavrakidisJonke2}
for details.} on a set of unitary operators $\hat U_i$ which commute with the 
$U_i$, i.e.
$$U_i\hat U_j=\hat U_j U_i.$$
The expression (\ref{gss}) for the general solution is reminiscent of the covariant coordinates which are used in the study of 
non-commutative gauge theories \cite{Madore:2000en}. They represent a non-commutative connection which gives rise to the gauge fields 
of the resulting theory. Finding these hatted operators in the present case is a straightforward task. One 
anticipates that they are related to the left-invariant vector fields of the nilmanifold; indeed it is found that 
$\hat U_i=e^{i\hat Y_i},$
where 
\be 
\hat Y_i=\hat p_i-\sfrac 12f_{ij}^{\ k}\hat x^j\hat p_k.
\ee
Then the algebra of the hatted operators, which is essentially the gauge algebra of the theory, 
is found to be
\bea 
\hat U_1\hat U_3&=&e^{-i\theta'_{13}}\hat U_3\hat U_1, \nonumber\\
\hat U_2\hat U_3&=&e^{-i\theta'_{23}}\hat U_3\hat U_2, \nonumber\\
\hat U_1\hat U_2&=&\hat U_3\hat U_2\hat U_1.
\eea
In the case of trivial gauge bundles, the primed parameters are related to the unprimed ones as 
$\theta'_{ij}=-\theta_{ij}.$
This shows that the $C^*$ algebra given by $\hat U$s, where the gauge fields $\hat{ A}^i$ live, is dual to the one of 
the operators $U^i$. In particular, it was shown in Ref. \cite{packer} that these $C^*$ algebras are Morita equivalent.

\section{The case of general fluxes}

One of our main motivations is the situation described by Jackiw in Ref. \cite{Jackiw}, 
exhibiting a very interesting interplay 
among a physical quantum-mechanical system and the mathematics of cocycles. Studying a quantum-mechanical particle 
moving under the influence of a magnetic field one may postulate non-canonical commutation relations of the form
\bea 
[\hat x^i,\hat x^j]&=&0, \nonumber\\
{[}\hat x^i,\hat p_j]&=&i\d^i_j, \nonumber\\
{[}\hat p_i,\hat p_j]&=&i\e_{ijk}B^k,
\eea
where $\hat p_i$ is the mechanical momentum and $B$ is the magnetic field. Then the Jacobiator of the momenta is computed as
\be
[\hat p_i,\hat p_j,\hat p_k]\equiv [\hat p_i,[\hat p_j,\hat p_k]]+\mbox{c.p.}=\mbox{div}B.
\ee
In the absence of magnetic sources the Jacobi identity is satisfied. However, this fails to happen when the magnetic field 
has non-vanishing divergence, in which case a 3-cocycle is present and therefore some  non-associativity too. This 
non-associativity is lifted and the cocycle is removed at the level of the unitary (translation) operators 
due to a Dirac quantization condition.  In the following, this structure serves as motivation to consider in the 
matrix model more general 
%$(  X,\tilde{  X})$ 
commutation relations than before and investigate their physical meaning.
Moreover, the special role of the phase space in the previous section is 
another motivation to go further and explore its full potential. Finally, 
according to Ref. \cite{cds}, the parameters $\theta_{ij}$ of non-commutativity
correspond to background values of the three-form of 11-dimensional 
supergravity or, equivalently, to the B field of 10-dimensional type 
IIA supergravity. It is then interesting to examine the way that fluxes 
can be described in this formalism.   

\subsection{Geometric flux revisited} 

Let us begin by revisiting the geometric flux case which was described in the 
previous section. Here we would like to pay more attention to the full 
phase space structure. Previously we considered the action of the unitary operators on the coordinate space but their action on the momentum space was 
not made explicit. Now we reconsider these operators, acting on functions 
on the phase space as
\bea \label{Ualg}
(U_1f)(\hat x^i;\hat p_i)&=&f(\hat x^1+1, \hat x^3+\sfrac 12 \hat x^2;
\hat p_2-\sfrac 12 \hat p_3)=e^{i\hat p_1}f( \hat x^3+\sfrac 12 \hat x^2;
\hat p_2-\sfrac 12 \hat p_3), \nonumber\\
(U_2f)(\hat x^i;\hat p_i)&=&f(\hat x^2+1,\hat x^3-\sfrac 12 \hat x^1;\hat p_1+
\sfrac 12 \hat p_3)=e^{i\hat p_2}f(\hat x^3-\sfrac 12 \hat x^1;\hat p_1+
\sfrac 12 \hat p_3),  \nonumber\\
(U_3f)(\hat x^i;\hat p_i)&=&f(\hat x^3+1)=e^{i\hat p_3}f(\hat x^i;\hat p_i),
\eea 
where we write explicitly only the directions which are subject to some transformation. These operators when acting on $ X^i$ reproduce 
the conditions (\ref{conditionstt3}) as they should. Moreover, from (\ref{Ualg}) we can deduce their action on the momentum space,  $\tilde{ X}_i\equiv\hat p_i$. The non-trivial relations are
\bea  
U_1\tilde{ X}_2 U_{-1}&=&\tilde{X}_2-\sfrac 12 \tilde{ X}_3,\nonumber\\
U_2\tilde{ X}_1 U_{-2}&=&\tilde{X}_1+\sfrac 12 \tilde{X}_3.
\eea
Since we are working in phase space, let us also consider the operators 
$\tilde U^i$ which are dual to the $U_i$ ones. The former are associated 
to translations in momentum space much like the latter are translation operators in position space\footnote{Loosely speaking, one can think of these operators as coming from (\ref{yy}) after exchanging $\partial/\partial  x^i$ with $\partial/\partial p_i$.}.  Therefore we consider
\bea 
(\tilde U^1f)(\hat x^i;\hat p_i)&=&f(\hat p_1+1)=e^{-i\hat x^1}f(\hat x^i;\hat p_i), \nonumber\\
(\tilde U^2f)(\hat x^i;\hat p_i)&=&f(\hat p_2+1)=e^{-i\hat x^2}f(\hat x^i;\hat p_i),\\
(\tilde U^3f)(\hat x^i;\hat p_i)&=&f(\hat p_1-\sfrac 12 \hat x^2,\hat p_2+\sfrac 12\hat x^1,\hat p_3+1)=e^{-i\hat x^3}f(\hat p_1-\sfrac 12 \hat x^2,\hat p_2+\sfrac 12\hat x^1).\nonumber
\eea 
These operators act trivially on the $ X^i$, however they have the following non-trivial action on $\tilde{ X}_i$:
\bea 
\tilde U^i\tilde{ X}_i \tilde U^{-i} &=& \tilde{ X}_i + 1, \nonumber\\
\tilde U^3\tilde{ X}_1 \tilde U^{-3} &=& \tilde{ X}_1 -\sfrac 12  X^2, \nonumber\\
\tilde U^3\tilde{ X}_2 \tilde U^{-3} &=& \tilde{ X}_2 +\sfrac 12  X^1.
\eea 
Setting for simplicity all the constant parameters $\theta_{ij}$ to zero, the 
extended algebra of the operators $U_i$ and $\tilde U^i$ has the following 
non-trivial relations:
\bea 
U_2U_1=U_3U_1U_2, \quad U_1\tilde U^3=\tilde U^2\tilde U^3U_1, \quad 
\tilde U^3U_2=\tilde U^1U_2\tilde U^3.
\eea 
This is an extension of the $C^{\ast}$ algebra of the previous section. 
Moreover, as previously, one has to determine the operators $\hat U_i$ and 
${\hat{\tilde U}^i}$ commuting with $U_i$ and $\tilde U^i$ and thus providing the dependence 
of the gauge fields. We are not going to write their explicit form here but it is instructive to write their commutation relations which read as
\bea 
\hat U_1\hat U_2\hat U_{-1}\hat U_{-2}=\hat U_3, \quad 
\hat{\tilde U}^3\hat U_1\hat{\tilde U}^{-3}\hat U_{-1}=\hat {\tilde U}^2, 
\quad \hat{\tilde U}^3\hat U_2\hat{\tilde U}^{-3}\hat U_{-2}=\hat {\tilde U}^{-1}.
\eea 
These relations represent the gauge algebra of the resulting theory and they 
are reminiscent of the gauge algebra of a 4-dimensional supergravity obtained 
from a dimensional reduction on a twisted torus \cite{km}.

\subsection{H flux}

Let us now perform the following transformation,
\be\label{dual1}
( X^3,\tilde{ X}_3) \rightarrow (\tilde{ X}_3,- X^3),
\ee
which preserves the Heisenberg relation $[ X^i,\tilde{ X}_i]=i$,
accompanied with the corresponding transformation on the unitary operators,
\be 
(U_3,\tilde{U}^3)\rightarrow (\tilde{U}^3,U_{-3}).
\ee
Computing the action of the unitary operators $U_i$ on $ X^i$ 
we find
\be \label{cch1}
U_i X^jU_{-i}= X^j+\d_i^j.
\ee
This set of conditions look exactly like the ones for a simple toroidal compactification of the model. However, there is 
more structure than just a torus. Indeed, writing
 $U_i=e^{iY_i}$ and $\tilde U^i=e^{i\tilde Y^i}$,
we find the relations
\be \label{uh}
U_iU_j=e^{-iF_{ijk}\tilde Y^k}U_jU_i,
\ee
where $F_{ijk}$ is fully antisymmetric. These relations exhibit a 3-cocycle. Therefore we interpret the situation as presence of some flux, which is obtained as a derivation 
of the cocycle. Such derivations are actually commutators and therefore in the present case the flux is obtained as 
\be \label{fh}
F_{ijk}=[{ Y}_{[i},[{ Y}_j,{ Y}_{k]}]].
\ee
This flux is interpreted as a NSNS H flux in the corresponding supergravity 
compactification. There are at least two arguments which support this interpretation. 
First, as advocated  in Ref. \cite{cds}, the cocycle appearing in Eq. (\ref{uh}) is 
related to the B field of type IIA supergravity.  
In supergravity compactifications  an 
H flux may be obtained from a B field with the general structure
$$B=\sfrac 13(x^1dx^2\wedge dx^3+x^2dx^3\wedge dx^1+ x^3dx^1\wedge dx^2).$$
The H flux is then obtained by
$$H_{ijk}=\partial_{[i}B_{jk]},$$
and the analogy to Eq. (\ref{fh}) is evident.

Another argument for the interpretation of the above situation as an H flux is the following. With the above data we may determine 
conditions from the action of $U$s on $\tilde{ X}$s. In particular we obtain that
\be \label{cch2}
U_i\tilde{ X}_jU_{-i}=\tilde{ X}_j-F_{ijk} X^k.
\ee
Moreover, the action of $\tilde U^i$ on $ X$s and 
$\tilde{ X}$s can be determined in a similar way and it is found to be
\bea \label{cch3}
\tilde{U}^i X^j\tilde{U}^{-i}&=& X^j, \\ \label{cch4}
\tilde{U}^i\tilde{ X}_j\tilde{U}^{-i}&=&\tilde{ X}_j+\d^i_j.
\eea
The tilded operators act trivially on $ X$s and as translations on $\tilde{ X}$s. Collecting together the conditions 
(\ref{cch1}), (\ref{cch2}), (\ref{cch3}), (\ref{cch4}), we see directly that these are the conditions that should be 
imposed for the compactification on a particular twisted doubled torus \cite{hulltdt,prezastdt}.
This twisted doubled torus is exactly the one associated 
with an H flux situation.

\subsection{Q flux}

A further transformation,
\be\label{dual2}
(X^2,\tilde{ X}_2) \rightarrow (\tilde{ X}_2,- X^2),\quad 
(U_2,\tilde{U}^2)\rightarrow (\tilde{U}^2,U_{-2}).
\ee
leads to the following  action of $U$s and $\tilde U$s on $ X$s and $\tilde{ X}$s:
\bea 
U_i X^i U_{-i} =  X^i + 1, \quad 
U_1 X^2 U_{-1} =  X^2 +\sfrac 12\tilde{  X}_3, \quad
U_1 X^3 U_{-1} =  X^3 -\sfrac 12\tilde{  X}_2, \label{fc1}
\eea
and
\bea
\tilde U^i\tilde{  X}_i \tilde U^{-i} &=& \tilde{  X}_i + 1, \nn\\
\tilde U^2{  X}^3 \tilde U^{-2} &=& {  X}^3 +\sfrac 12   X^1, \quad
\tilde U^3{  X}^2 \tilde U^{-3} = {  X}^2 -\sfrac 12   X^1, \nn\\
\tilde U^2\tilde{  X}_1\tilde U^{-2} &=& \tilde{  X}_1+\sfrac 12\tilde{  X}_3, \quad
\tilde U^3\tilde{  X}_1\tilde U^{-3} = \tilde{  X}_1-\sfrac 12\tilde{  X}_2.
\eea
The corresponding operator algebra is specified by the non-trivial relations
\be 
U_i\tilde U^j=e^{F_i^{jk} Y_k}\tilde U^jU_i, \quad
\tilde U^i\tilde U^j=e^{-F^{ij}_k\tilde Y^k}\tilde U^j\tilde U^i.
\ee 
In the same spirit as before, the above compactification conditions 
correspond to the global identifications for the coordinates of a twisted doubled torus associated to non-geometric $Q$
flux in Refs. \cite{hulltdt,prezastdt}. This, along with a further argument that we discuss later, suggests that the  generalised 
flux $F^{ij}_k$ should be identified with a $Q^{ij}_k$ non-geometric flux.

It should be noted that in the present case 
 the sector of $  X$s does not close in itself and therefore a 
truncation of  the compactification to the sector of $  X$s and $U$s is not plausible. We discuss how this situation should be 
interpreted in the following section.

\subsection{R flux}

As before, let us consider the last remaining possible transformation,
\be\label{dual3}
(  X^1,\tilde{  X}_1) \rightarrow (\tilde{  X}_1,-  X^1),
\quad (U_1,\tilde{U}^1)\rightarrow (\tilde{U}^1,U_{-1}).\ee 
which preserves the Heisenberg relation.
The action of $U$s and $\tilde U$s on $  X$s and $\tilde{  X}$s is now given as
\bea 
U_i  X^jU_{-i}&=&  X^j+\d_i^j, \nonumber\\
U_i\tilde{  X}_jU_{-i}&=&\tilde{  X}_j, \nonumber\\
\tilde{U}^i  X^j\tilde{U}^{-i}&=&  X^j+F^{ijk}\tilde{  X}_k,\nonumber\\
\tilde{U}^i\tilde{  X}_j\tilde{U}^{-i}&=&\tilde{  X}_j+\d^i_j,
\eea
where $F^{ijk}$ is fully antisymmetric.
These relations correspond to the identification conditions of a twisted doubled torus with R flux as in Refs. \cite{hulltdt,prezastdt}.
The non-trivial relations of the operator algebra are
\be 
\tilde U^i\tilde U^j=e^{-F^{ijk}Y_k}\tilde U^j\tilde U^i.
\ee 
Further arguments 
related to this case are discussed below.

\section{Interpretation and relations to other frameworks}

\paragraph{Relation to  generalised fluxes in Double Field Theory.}

As we mentioned in the introduction, there is a variety of frameworks addressing the issues of non-geometry and non-geometric 
fluxes. A mathematical object which plays a central role in most of them is the  generalised metric. Here we are not going to 
provide any details on  generalised geometry due to lack of space. Instead we collect just the necessary 
ideas and formulas for our interpretation and 
refer the reader to other self-contained texts.

The  generalised metric collects a d-dimensional metric $g$ and the 2-form field $B$
in a 2d-dimensional matrix which has the form
\be
  H = \left(\begin{array}{cc} g-Bg^{-1}B  & Bg^{-1} \\ 
-g^{-1}B & g^{-1} \end{array}\right).
\ee
This  generalised metric is obtained from a  generalised vielbein $  E$ as $  H=  E^{T}  E$, where 
\be
  E = \left(\begin{array}{cc}  e & 0 \\ 
-e^{-T}B & e^{-T} \end{array}\right),
\ee
and $e$ is the standard vielbein out of which the metric $g$ is constructed. 
The above parametrization of the  generalised metric is not unique; assuming that $B\beta=0$ and 
writing  the  generalised vielbein as
\be
  E = \left(\begin{array}{cc} e  & e\beta \\ 
-e^{-T}B & e^{-T} \end{array}\right),
\ee
leads to the following parametrization of the generalised metric 
\be\label{hgen}
  H = \left(\begin{array}{cc} g-Bg^{-1}B  & Bg^{-1}+g\beta \\ 
-g^{-1}B-\beta g & g^{-1}-\beta g \beta \end{array}\right).
\ee
In the latter expressions $\beta=\beta^{ij}\partial_i\wedge \partial_j$ is a bivector, 
which appears naturally in generalised geometry. Moreover, it is very useful in the study of non-geometric cases 
in this framework \cite{Grana:2008yw}. In order to relate this discussion with the formalism that was presented in the previous section 
we would like to mention the case of generalised Scherk-Schwarz compactifications of double field theory. These 
were studied in Refs. \cite{DF1,DF2,Geissbuhler:2013uka,Grana:2012rr,Berman:2012uy}
and led to expressions of generalised geometric and non-geometric fluxes in terms of 
the vielbein $e$, the 2-form $B$ and the bivector $\beta$. Although in the general case these expressions are 
complicated, for the simple case of the Heisenberg nilmanifold they simplify significantly. Then the non-geometric 
fluxes $Q$ and $R$ are obtained as the derivative and the dual derivative of the bivector respectively, where the 
dual derivative is with respect to the dual coordinates of double field theory. We summarise these expressions in the 
following 
table, where we also show how the corresponding generalised fluxes are obtained in our formalism.

\begin{center}
 \begin{tabular}{l*{4}{c}r}
 generalised Flux       & Matrix Model       & DFT  \\ 
\hline \vspace{5pt}
$H_{ijk}$  & $[{  Y}_{[i},[{  Y}_j,{  Y}_{k]}]]$ &  $\partial_{[i}B_{jk]}$  \\ \vspace{5pt}
$f_{ij}^k$         &  $[\tilde{  Y}^{k},[{  Y}_i,{  Y}_{j}]]$ & $\tilde\partial^kB_{ij}$   \\ \vspace{5pt}
$Q^{ij}_k$ & $[{  Y}_{k},[\tilde{  Y}^i,\tilde{  Y}^{j}]]$ & $\partial_k\beta^{ij}$ \\ \vspace{5pt}
$R^{ijk}$         & $[\tilde{  Y}^{[i},[\tilde{  Y}^j,\tilde{  Y}^{k]}]]$ & $\tilde\partial^{[i}\beta^{jk]}$   \\ \hline
\end{tabular} 
\end{center}

We observe that the structure of these relations in the matrix model and in double field theory agree and they are in  
direct correspondence. Let us remember that according to Ref. \cite{cds} the quantity $[{  Y}, {  Y}]$ 
corresponds to the B field in supergravity. The first two lines of the above table are reminiscent of this relation. 
What is more, the last two lines of the table indicate a further relation, which is that the quantity $[\tilde{  Y}, \tilde{  Y}]$ 
corresponds to the bivector $\beta$ in the  generalised geometry approach to supergravity.

We would like to point out that in richer situations, more complex that the Heisenberg nilmanifold, the above simplification 
of the form of the  generalised fluxes is not always possible. In other words, there exist higher-dimensional nilmanifolds 
(six-dimensional in particular) where in certain duality frames the  generalised metric cannot be written in terms of $g$ and $B$ or 
$g$ and $\beta$ only but the general structure of Eq. (\ref{hgen}) is necessary. We shall examine such cases
in a forthcoming publication.

\paragraph{Phase space and non-geometric matrix compactifications.}

It should be obvious from  the previous sections that phase space plays an important role in the present formalism. 
The pair of $(  X,\tilde{  X})$ is reminiscent of coordinates and canonical momenta in phase space. Moreover, as it was noticed in section 
3, while for the cases with $H$ and $f$ flux it is possible to project the compactification to the $  X$-sector, this 
is no longer true for the cases with $Q$ and $R$ flux. The interpretation of this situation is that in the latter cases 
the meaningful projection is on the $\tilde{  X}$-sector. In quantum-mechanical terms this is like shifting to the 
momentum representation. The picture then is that in the matrix model formalism geometric compactifications with $H$ or 
$f$ flux are well-defined in position space ($X$-sector), while the non-geometric ones with $Q$ or $R$ flux are well-defined 
in momentum space ($\tilde{X}$-sector). A similar correspondence follows for the non-commutativity parameters 
$i\tilde\theta^{ij}=[  \tilde{Y}^i,  \tilde{Y}^j]$ and $i\theta_{ij}=[{  Y}_i,{  Y}_j]$. In position space there 
are compactifications with non-constant $\theta_{ij}$ corresponding to the geometric cases and in momentum space there 
are compactifications with non-constant $\tilde\theta^{ij}$ which correspond to the non-geometric cases.    A similar result 
in the context of  generalised complex geometry was obtained in Ref. \cite{Andriot1}. 

\paragraph{Flux Quantization.}

Another property of the models that we discuss here, stemming from their quantum-mechanical nature, is related to the 
quantization of flux. It is known that in quantum-mechanical problems with non-constant fields and 
non-canonical commutation relations there are Dirac quantization conditions which follow from consistency requirements 
\cite{Jackiw}. This is essentially due to the fact that at the level of finite translations any 3-cocycle, which is also 
a signature of failure of the Jacobi identity, should be removed. In the present formalism and taking as an example the 
case of $H$ flux, the translation operators 
are the $U_i$s and they satisfy the relation
\be \label{una}
U^i(U^jU^k)=e^{\frac i2 H^{ijk}}(U^iU^j)U^k.
\ee 
This cocycle condition is rendered trivial by the quantization condition
\be
H=4\pi n, \quad n\in\Z.
\ee
Needless to say that this is in accord with string theory, where all charges have to be quantized.

\vspace{20pt}
\paragraph{Acknowledgements.} We would like to acknowledge the very pleasant atmosphere of the Corfu Summer Institute 
which offered the basis for vivid discussions in a wonderful environment. We would particularly like to thank 
I. Florakis, D. Giataganas, C. Hull, N. Irges, M. Schmitz, P. Schupp and R. Szabo for discussions.
A. C. acknowledges partial support from the grant DFG  LE 838/13.

\end{document}